# Effect of near-earth thunderstorms electric field on the intensity of ground cosmic ray positrons/electrons in Tibet

X. X. Zhou*, X. J. Wang, D. H. Huang, H. Y. Jia

( *School of Physical Science and Technology, Southwest Jiaotong University, Chengdu 610031, China* )

**Abstract**: Monte Carlo simulations are performed to study the correlation between the ground cosmic ray intensity and near-earth thunderstorms electric field at YBJ (4300 m a.s.l., Tibet, China). The variations of the secondary cosmic ray intensity are found to be highly dependent on the strength and polarity of the electric field. In negative fields and in positive fields greater than 600 V/cm, the total number of ground comic ray positrons and electrons increases with increasing electric field strength. And these values increase more obviously when involving a shower with lower primary energy or a higher zenith angle. While in positive fields ranging from 0 to 600 V/cm, the total number of ground comic ray positrons and electrons declines and the amplitude is up to 3.1% for vertical showers. A decrease of intensity occurs in inclined showers within the range of 0−500 V/cm, which is accompanied by smaller amplitudes. In this paper, the intensity changes are discussed, especially concerning the decreases in positive electric fields. Our simulation results are in good agreement with ground-based experimental results obtained from ARGO-YBJ and the Carpet air shower array. These results could be helpful in understanding the acceleration mechanisms of secondary charged particles caused by an atmospheric electric field.

**Keywords**: near-earth thunderstorms electric field; cosmic rays; extensive air showers; Monte Carlo simulations; ARGO-YBJ

**PACS**: 92.60.Pw; 92.60.Ta; 52.80.Mg

## 1. Introduction

The effect of thunderstorms electric field on the development of cosmic ray air showers, especially on the intensity of secondary cosmic rays, is one of the hottest topics in high-energy atmospheric physics. During thunderstorms, the maximum strength of electric fields has been found in the range of 850−1300 V/cm [1], or even up to 2000 V/cm [2, 3]. In such strong fields, by accelerating or decelerating the charged particles in extensive air showers, the intensity of secondary cosmic rays could be influenced. It was first suggested by Wilson [4] in 1924 that the strong electric field during thunderstorms might result in an observable effect on a secondary electron, which has tiny mass. In 1992, Gurevich et al. [5] suggested an avalanche-type increase of the number of runaway electrons could lead to a new type of electric breakdown of gases in the atmosphere. They developed the theory of runaway breakdown (RB), now mostly referred to as relativistic runaway electron avalanche (RREA) [6]. Marshall et al. [3, 7], Dwyer [8] and

* Corresponding author. E-mail: zhouxx@swjtu.edu.cn (X. X. Zhou)

Symbalisty et al. [9] studied the strength of threshold field necessary for an avalanche to occur, which is dependent on the altitude. They found that the value was about 2800 V/cm at sea level.

For decades, many scientists have carried out a wide range of ground-based experiments to detect the thunderstorm ground enhancement (TGE), a new high-energy phenomenon originating in the terrestrial atmosphere, trying to find high-energy electrons accelerated by thunderstorms electric field or high-energy photons radiated by bremsstrahlung. The intensity enhancements of ground cosmic rays have been detected by high altitude experiments, such as the Carpet air shower array [10, 11], EAS-TOP [12], ASEC [13-17] and ASγ [18]. Their results indicated that the increases were associated with the electric field and the RREA process could be responsible for huge TGEs. Tsuchiya et al. [18-20] and Torii et al. [21-22] provided clear evidence that strong electric fields can accelerate electrons beyond a few tens of MeV.

It is well known that the strong electric discharges associated with thunderstorms can produce terrestrial gamma-ray flash (TGF). For years, thousands of TGFs have been detected by satellite-based experiments, such as AGILE [23] and GBM-Fermi [24]. The lightning initiation and correlations with thunderstorms have also been studied in details [16, 25-29].

To discover more valuable information, a few simulations have been done to study the intensity and energy changes of secondary particles during thunderstorms [17, 30-32]. Buitink et al. [33] have modified the CORSIKA code and performed simulations to calculate the effect of an electric field on the development of proton showers with energies more than $10^{16}$ eV. Their results found that the RREA might occur at high altitudes.

From the experimental observations and simulation results above, it seems that these enhancements of secondary particles during thunderstorms are correlated with the electric field and the RREA will occur under certain conditions. However, the acceleration mechanisms of secondary charged particles caused by atmospheric electric field still remain unresolved.

In 2011, the AGILE team found that the TGF emission above 10 MeV had a significant power-law spectrum with energies up to 100 MeV [34]. These results posed a big challenge for the widely accepted TGF model based on the RREA mechanism.

It is clear now that intensity decreases for the hard component of cosmic rays are associated with thunderstorms electric field. Chilingarian et al. [15] found a deficit of ~6.0% in the flux of muons with energies greater than 200 MeV during thunderstorms. Alexeenko et al. [11] studied the effects of thunderstorms electric field on the soft and hard components of cosmic rays separately. The net effect is a decreasing intensity for the hard component (muons) and an increasing intensity for the soft component (electrons). Interestingly, a negative correlation of variations between the electric field and the soft component intensity was reported in the same paper. That is to say, the intensity of the soft component decreases in a certain range of positive fields. The study suggested that the reason for this decreasing phenomenon was the poor separation of the soft and hard components. Is the soft component intensity decrease related to thunderstorms electric field or poor separation of the components?

The intensity changes of ground cosmic rays were detected by the ARGO-YBJ experiment (located at YBJ, Tibet, China). These changes are related to the multiplicity ($n$) of charged particles (mostly for positrons and electrons) in scaler mode. An increase of the single particle counting rate occurs accompanied with thunderstorms electric field for channel $n = 1$ or $n = 2$. But if $n = 3$ or $n \geq 4$, the counting rate does not obviously change or may even decline [35, 36] (see in Fig.1). The intensity decrease cannot be explained by the RREA mechanism. Are these soft components decreasing phenomena associated with thunderstorms electric field? Moreover, what is the acceleration mechanism for them?

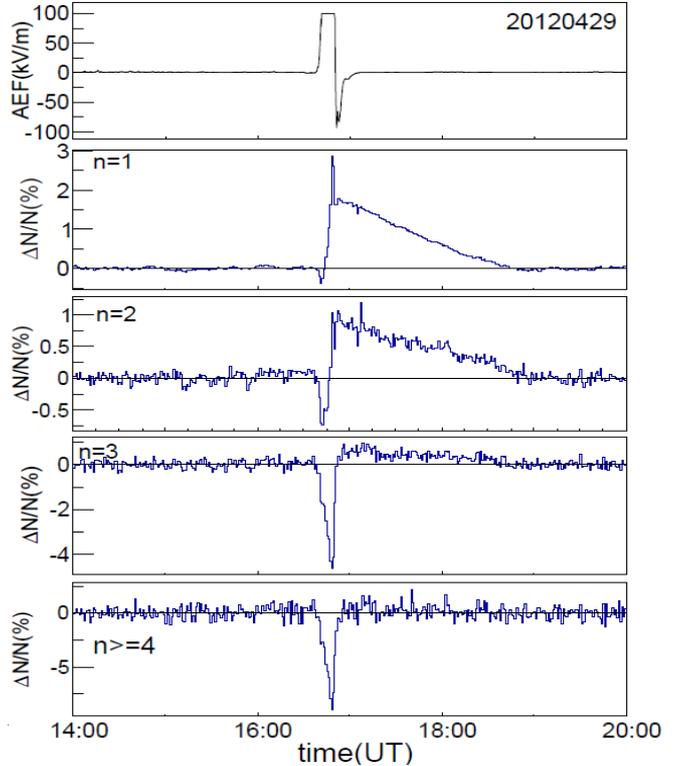

Fig. 1 Thunderstorm event detected on April 29, 2012. From top to bottom, the value of atmospheric electric field (AEF), the percent variations of counting rate for $n = 1$, $n = 2$, $n = 3$ and $n \geq 4$ are shown [36].

Because of the unknown strength and structure of thunderstorms electric field, there are numerous problems regarding the processes of high-energy particle interactions in the atmosphere that remain unsolved to this today. In order to learn more about the acceleration mechanism and the intensity change, more theoretical, experimental and careful simulation results are needed.

In this work, we perform Monte Carlo simulations by using CORSIKA to study the effect of a near-earth electric field on the intensity of ground cosmic ray positrons/electrons at YBJ. Using these simulations, we then try to analyze the cause of the decreasing phenomena for soft components. This paper is organized as follows: The simulation parameters are introduced in Section 2. The simulation results of vertical and inclined proton showers with several different primary energies are shown in Section 3. The discussions of intensity changes, especially those involving decreasing phenomena in positive fields, are presented in Section 4. The conclusions are given in Section 5.

**2. Simulation parameters**

CORSIKA (COsmic Ray SImulations for KAscade) is a detailed Monte Carlo program used to study the evolution and properties of extensive air showers in the atmosphere [37]. In our simulation work, we use the code of CORSIKA 7.3700, its subroutine ELECTR has been extended to account for the effect of atmospheric electric fields on the transport of electromagnetic particles. The extension follows the programming procedure *emf_macros.mortran*, which was developed by Bielajew [38]. The selected hadronic interaction model is QGSJETII-04 for high energy and GHEISHA for low energy.

Previous studies have shown that the atmospheric electric field distributed roughly within the altitude scope of 4−12 km during thunderstorms [39]. Because charged secondary particles will lose their energies quickly through radiation and ionization, the effect on the intensity of charged particles can be neglected in the electric field, which is far from detectors. In our simulations, the electric field length is 2000 m, from an altitude of 6300 m to 4300 m (corresponding to the atmospheric depth 484−606 g/cm$^2$). It has been found that the strength of near-earth thunderstorms electric field at YBJ is mostly within 1000 V/cm [40]. In our work, the uniform electric field ranges from -1000 to 1000 V/cm. Here, we define the positive electric field as one that accelerates positrons downward in the direction of the earth.

According to the energy threshold of the ARGO-YBJ detector, which is a few tens of GeV in scaler mode and a few hundreds of GeV in shower mode [41], proton showers with energies of 30, 100 and 770 GeV are chosen as the primary particles in this work.

Since positrons and electrons predominate in the secondary charged particles of cosmic rays, and the hadronic and muonic parts of the shower are hardly affected, the effects of the electric field on positrons and electrons are properly taken into account in our work. In view of the acceleration of the field, the energy cutoff is set to 0.1 MeV, below which value positrons and electrons are discarded from the simulation.

**3. Simulation results**

When a primary cosmic ray enters the atmosphere, it will produce a large number of secondary particles via the hadron and electromagnetic cascades. These particles are distributed in a range many kilometers wide. This phenomenon is called extensive air shower (EAS). The total number of secondary particles, which are produced in an EAS at a particular level in the atmosphere, is called the shower size. In this paper, we only consider the effect of an electric field on positrons and electrons. The shower size, namely multiplicity (*n*), is defined as the total number of positrons and electrons.

**3.1 Vertical showers with primary energy 100 GeV**

Considering the fluctuations from shower to shower, we generate $2 \times 10^6$ vertical proton showers with primary energy 100 GeV using CORSIKA. The electric fields are chosen as a series of values in the range of -1000−1000 V/cm. The correlations between the number of positrons/electrons and the near-earth electric field are simulated. Fig. 2 shows the percent change of the average number of positrons, electrons and the sum of both in different electric fields at YBJ.

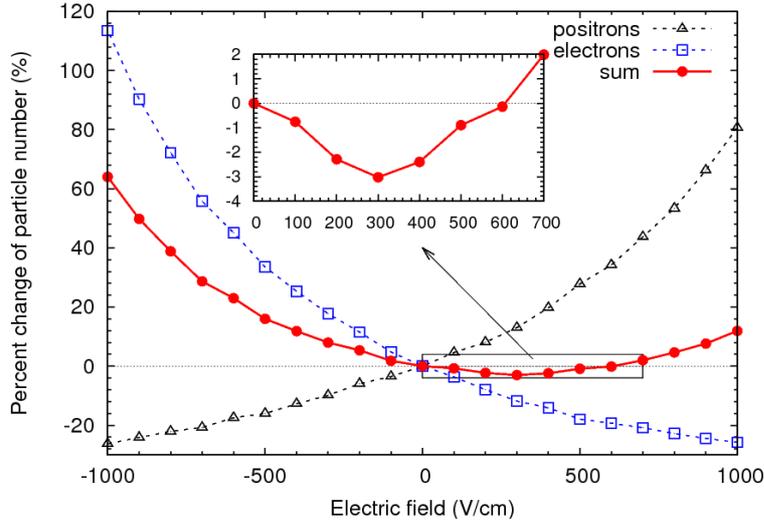

Fig. 2 Percent change of the average number of positrons, electrons and sum of both as a function of electric field at YBJ (The illustration is the enlarged view of the total number in reducing range).

As shown in Fig. 2, when the electric field is negative (accelerating electrons), the number of electrons ($N_{e-}$) increases, while the number of positrons ($N_{e+}$) decreases. The shower size (total number of positrons and electrons) increases as the electric field strength increases, and the amplitude enhancement is up to 66.5% in an electric field of -1000 V/cm. When the field is positive (accelerating positrons), the number of electrons decreases, while the number of positrons increases. In positive fields greater than 600 V/cm, the shower size increases as the field strength increases, and the amplitude is lower than 14.3% in an electric field of 1000 V/cm. The amplitude of the enhancement is much lower than that in a negative field of the same strength. In the range of 0−600 V/cm, the shower size declines and the maximum amplitude is about 3.1%. From Fig. 2, the shower size is enlarged in all negative fields and positive fields greater than 600 V/cm, while in positive fields less than 600 V/cm, the shower size is reduced.

Fig. 3 shows the percent change of the average shower size as a function of atmospheric depth in several positive electric fields. When the field is switched on with an altitude of 6300 m (~484 g/cm$^2$), the shower size drops steeply and the degree of decline reaches up to 3.9% , which is mostly due to more low energy electrons losing their energies to be below the detection threshold in positive fields. Soon, it increases with an increasing atmospheric depth. At YBJ, the shower size is increased in 700 V/cm. While the shower size does not show any significant change in 600 V/cm, it is obviously decreased in 400 V/cm.

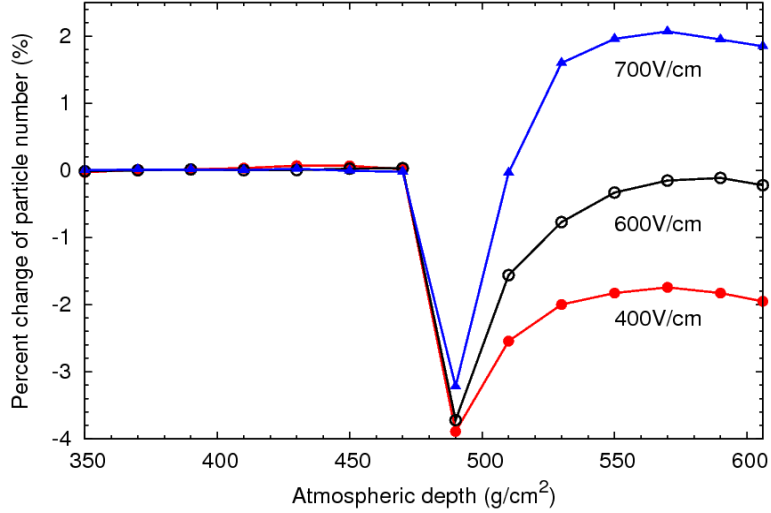

Fig. 3 Percent change of the total number of positrons and electrons as a function of atmospheric depth in 400, 600 and 700 V/cm.

As shown in Fig. 4, the correlations between the changes of number for different multiplicities and electric fields are plotted to compare with the observations of the ARGO-YBJ experiment. We find that the changes are related to the strength and polarity of electric fields. In negative fields and in positive fields greater than 600 V/cm, the number of multiplicity $n \leq 3$ decreased, and the number of multiplicity $n > 3$ increased. However, the situation is reversed in positive fields less than 600 V/cm; namely, the number is increased for $n \leq 3$ and decreased for $n > 3$. During thunderstorms, ARGO-YBJ detected an increase in the channel with lower multiplicity and a decrease in the channel with higher multiplicity. Our simulation results in 0−600 V/cm are in good agreement with the observations from the ARGO-YBJ experiment.

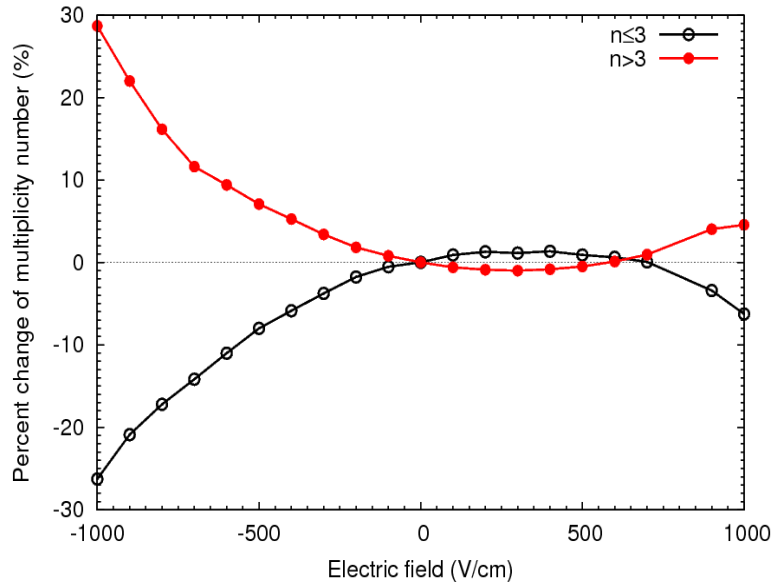

Fig. 4 Percent change of the number for different multiplicities as a function of electric field for $n \leq 3$ and $n > 3$.

Considering the responses of the ARGO-YBJ detector, we analyzed secondary positrons/electrons with energies more than 3 MeV. Similar results were obtained, as shown in Fig. 4.

**3.2 Vertical showers with primary energies of 30, 100 and 770 GeV**

This work simulated vertical proton showers with primary energies of 30, 100 and 770 GeV. In Fig. 5, the percent change of the average shower size is plotted as a function of electric field for several shower energies at YBJ. The black solid circle data points correspond to a primary energy of 30 GeV, the red hollow circle points for 100 GeV and the blue solid triangle points for 770 GeV.

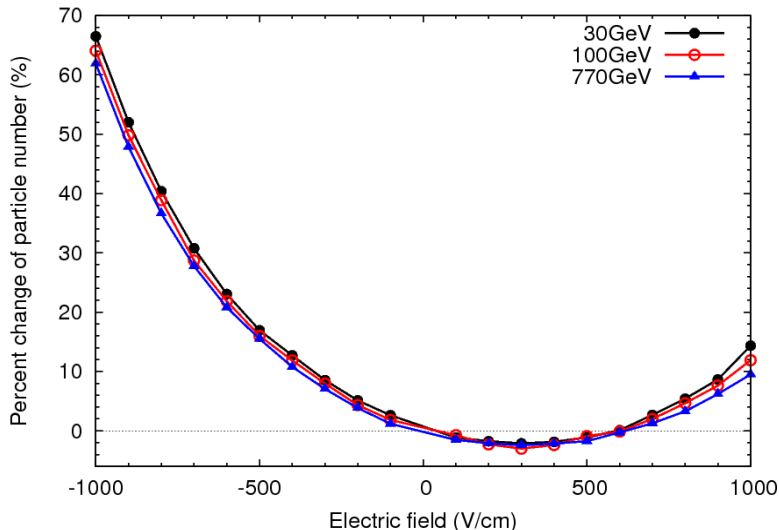

Fig. 5 Percent change of the total number of positrons and electrons as a function of electric field for primary energies of 30, 100 and 770 GeV at YBJ.

As we can see from Fig. 5, in negative fields or in positive fields greater than 600 V/cm, the intensities increase and the amplitude enhancements are larger for showers with lower primary energies. In the 0−600 V/cm range, obvious declines of the total number can be seen for all these different primary energy showers. The variation tendencies for 30, 100 and 770 GeV are almost the same.

**3.3 Inclined showers with primary energy 100 GeV**

Fig. 6 shows the results of inclined (with zenith angles of 30 and 60 degrees) proton showers with a primary energy of 100 GeV in different fields at YBJ. The red hollow circle points correspond to showers with a 30 degree angle and the blue solid triangle points for 60 degrees. Here, vertical showers (black solid circle points) are plotted for reference. As shown in the figure, decreasing phenomena occur in positive fields less than 500 V/cm and the maximum amplitude is about 1.7%, which is smaller than that of the vertical showers. In negative fields and in positive fields greater than 500 V/cm, intensities increase with an increasing field strength and zenith angle. We can see that the effects of positive/negative fields on positrons/electrons of inclined showers are stronger, especially for those with a 60 degree angle. For other results (such as the longitudinal development) concerning inclined showers, please refer to the report by Buitink et al. [33].

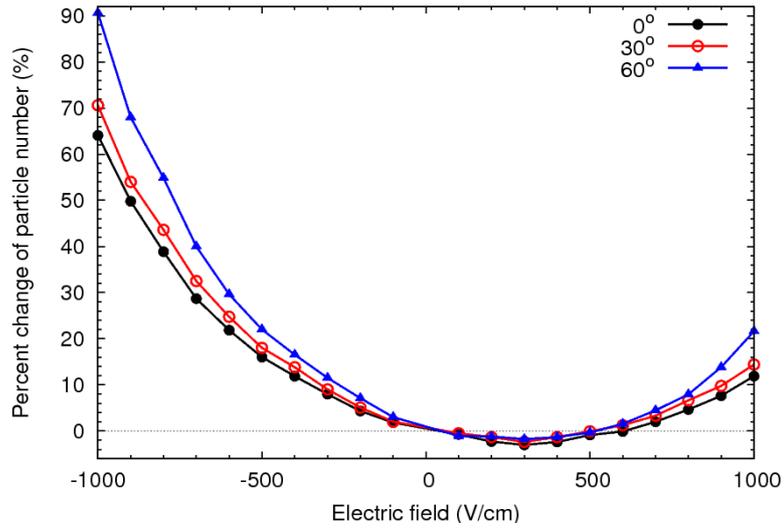

Fig. 6 Percent change of the total number of electrons and positrons as a function of electric field for vertical and inclined showers at YBJ.

**4. Discussions**

From our simulation results, we can see that intensity changes of ground cosmic ray positrons/electrons are associated with near-earth thunderstorms electric fields. The degree of intensity increase or decrease is dependent on the polarity and strength of the particular electric field, as well as the primary energy and zenith angle of the shower. We will discuss the intensity changes especially for decreasing phenomena from the effects as follows: the ratio of electrons to positrons, the energy of positrons and electrons, and the energy and the zenith angle of the primary shower. In the following discussions, if not otherwise specified, all results are for vertical showers with a primary energy of 100 GeV.

**4.1 The ratio of electrons to positrons**

It is well known that the number of positrons is less than that of electrons in cosmic rays for the asymmetry of mechanism, including Compton scattering, positron annihilation and photo-electric effects. Fig. 7 shows the percent of positrons and electrons in the total number as a function of atmospheric depth in absence of an electric field. From our simulation results, we can clearly see that the percentage of positrons decreases as the atmospheric depth increases, while it increases for electrons. For example, at 100 g/cm$^2$, the number of electrons is about 1.5 times of that of positrons; at YBJ (606 g/cm$^2$), the value is up to more than 1.8. The reason is mostly that the Compton scattering and photo-electric effects will increase as the atmospheric depth increases.

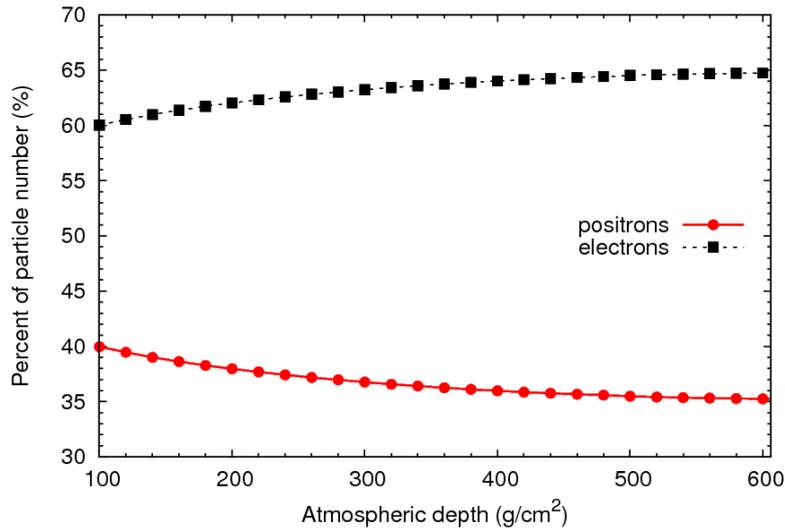

Fig.7 Percent of electrons and positrons in the total number as a function of atmospheric depth in *E*=0.

Fig. 8 shows ratios of electrons to positrons ($N_{e-}/N_{e+}$) in different electric fields at YBJ. In negative fields, the number of electrons is much greater than that of positrons, the ratio is more than 1.8 and the value increases with as the field strength increases. In positive fields, positrons are accelerated and low energy positrons gain sufficient energy to be above the detection threshold, and the ratio declines as the field strength increases. In positive fields greater than 600 V/cm, the ratio is less than 1.0 (< 0.75 in 1000 V/cm), which means that positrons outnumber electrons. But in positive fields less than 600 V/cm, the ratio is larger than 1.0, which means that the number of positrons is still less than that of electrons. The number of electrons decreases faster than the number of positrons increase in these fields. This factor ultimately leads to a deficit in total number of positrons and electrons in positive fields less than 600 V/cm.

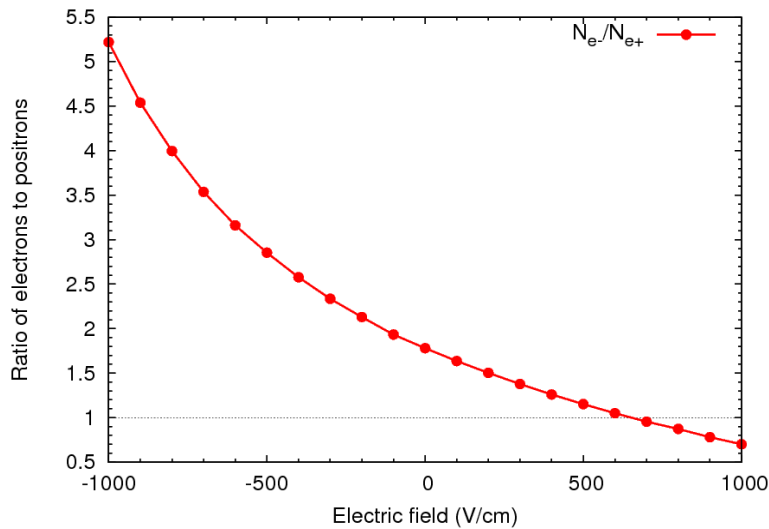

Fig. 8: Ratios of electrons to positrons as a function of electric field at YBJ.

## 4.2 The energy of positrons and electrons

Fig. 9 shows energy distributions of positrons and electrons at 6300m in absence of an electric field. In low energy regions, between 1−12 MeV, the proportion of electrons, which is the ratio of electrons in a certain energy region to the total number of electrons, is much larger than that of positrons. But the situation is reversed in higher energy regions. Positrons with energies above 12 MeV become more dominant than that of electrons. For example, there are 71.7% positrons but only 51.7% electrons with energies more than 12 MeV.

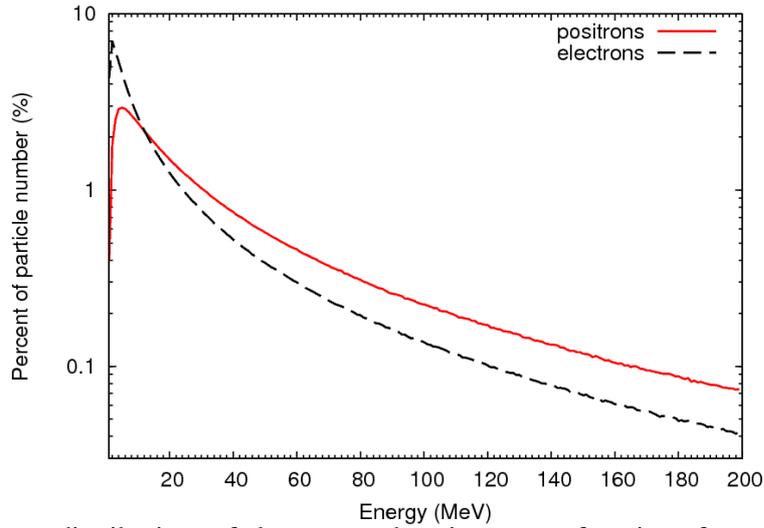

Fig. 9 Percent distributions of electrons and positrons as a function of energy at 6300 m.

It is well known that the slowing-down force of positron/electron in the atmosphere varies with its energy. According to the theory of Bethe [42], if the energy of positron/electron is more than ~1 MeV [43], the drag force increases as the energy increases. In electric field, the equilibrium energy ($U$) [33] of positron/electron can be expressed by $U(X) = \dfrac{qEZ_0 X_0}{X}$, where $X$ is the atmospheric depth with unit g/cm$^2$, $X_0$ (~36.7 g/cm$^2$) is the radiation length for electron/positron in air, $Z_0$ (~8.4 km) is the scale height, and $E$ is the electric field. Particles below equilibrium energy are accelerated. Radiation losses are dominated for particles above this energy. At the same altitude, the equilibrium energy is higher in stronger fields. For example, at an altitude of 6300 m, the equilibrium energy is about 19.3 MeV in 300 V/cm and 64.2 MeV in 1000 V/cm. From Fig. 9, it can be seen that there are about 59.8% electrons and only 40.9% positrons, which can be accelerated in the field strength of 300 V/cm at 6300 m. It shows that electrons with lower energies are easier to be affected in the same strength field. This factor also leads to a deficit in the total number of positrons and electrons in positive fields.

Fig. 10 shows the number change of positrons, electrons and the sum of both as a function of energy in 300 V/cm at YBJ. When energies are below equilibrium energy, which is less than 19.3 MeV at an altitude lower than 6300 m in 300 V/cm, the number of positrons and electrons changes noticeably. In low energy regions, the increased number of positrons is clearly smaller than the decreased number of electrons. In energies above 12 MeV, the proportions of positrons are larger than that of electrons from Fig. 9, and the total number increases insignificantly. As a result, a decline occurs in 300 V/cm.

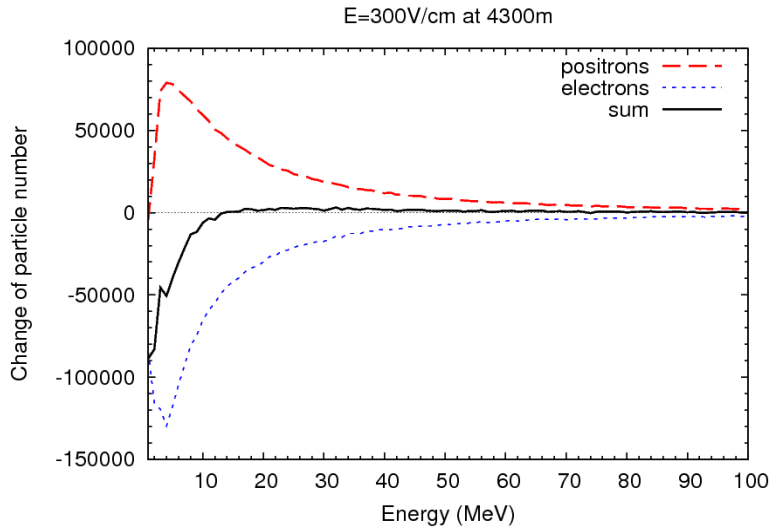

Fig. 10 Number changes of electrons, positrons and sum of both as a function of energy at YBJ in 300 V/cm.

Fig. 11 shows the energy distributions of positrons and electrons at YBJ in electric fields of 300 and 1000 V/cm. Solid lines correspond to positrons and dashed lines correspond to electrons. The red thin lines stand for 300 V/cm and blue bold lines stand for 1000 V/cm.

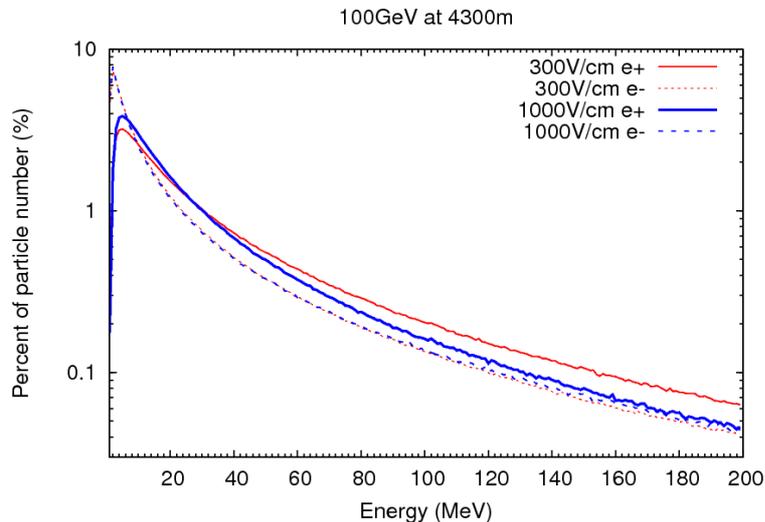

Fig. 11 Percent distributions of positrons and electrons as a function of energy at YBJ in 300 and 1000 V/cm, respectively.

As seen in Fig. 11, there are still more positrons with higher energies than that of electrons, and the energies of positrons in 300 V/cm are clearly higher than that in 1000 V/cm. That is to say, more low energy positrons are generated by pair production in 1000 V/cm. Considering the ratio of electrons to positrons and the equilibrium energy, 18.0% of the total particles can be accelerated and 35.4% can be decelerated in 300 V/cm; while in 1000 V/cm, the values vary from 47.5% to 36.6%. It is easy to understand why a decrease occurs in 300 V/cm while an increase occurs in 1000 V/cm.

From the discussions above, we know that the number of electrons is greater than that of positrons and there are higher proportions of positrons with larger energies than that of electrons. In certain positive fields, the increase of positrons cannot be compensated by the decrease of electrons, and so an obvious decline in the total number of positrons and electrons occurs.

**4.3 Primary energy and zenith angle of the shower**

In order to understand more about the relation between the intensity changes and the primary energies of showers, we analyze the energy and number distributions of positrons and electrons. Fig. 12 shows the energy distributions for showers with 30, 100 and 770 GeV at an altitude of 6300 m. We can see there are more positrons and electrons with larger energies for higher primary energy showers. Fig. 13 shows the ratios of electrons to positrons as a function of atmospheric depth in the absence of an electric field. There are more positrons for higher primary energy showers. The reason is mostly that the effect of pair production exceeds the Compton scattering for higher energy showers.

For higher primary energy showers, because of higher energies of positrons/electrons, the effect of the electric field on positrons/electrons is weaker, and the increase in shower size is smaller in negative fields and in positive fields greater than 600 V/cm. In positive fields less than 600 V/cm, the size of the shower with higher primary energy could be easily enlarged for the lower ratio of electrons to positrons, but it is more difficult to accelerate the positrons with higher energies. Because of the contradictory effects of these two factors, the differences of intensity changes are not obvious in the 0-600 V/cm range.

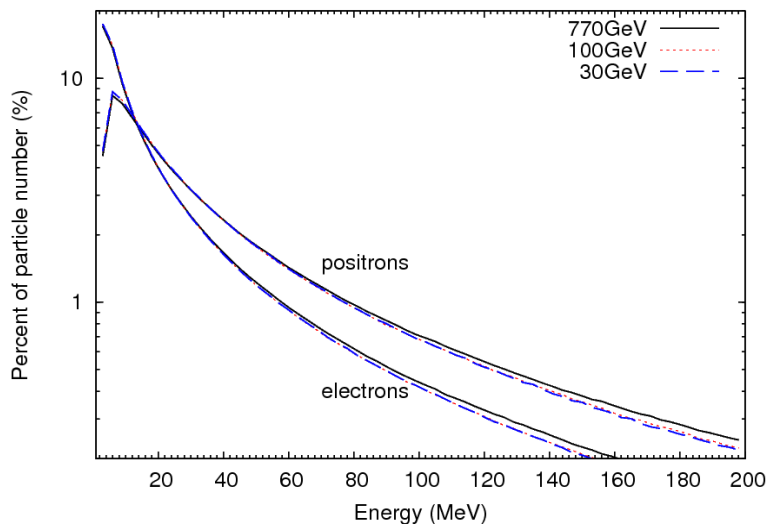

Fig. 12 Percent distributions of electrons and positrons as a function of energy for different primary energies at 6300 m

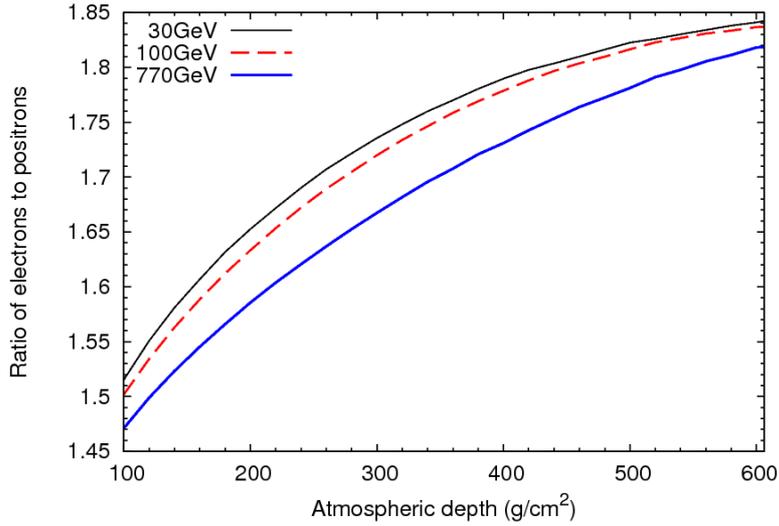

Fig. 13 Ratios of electrons to positrons as a function of atmospheric depth for several primary energies in *E* = 0.

Fig. 14 and Fig. 15 show the energy distributions and ratios of electrons to positrons for vertical and inclined showers. As for the inclined shower, we only show the results of those with a 60 degree zenith angle. From the two figures, we can see that the energies and ratios of inclined showers are lower than vertical showers at YBJ. Furthermore, because of the larger atmospheric length for inclined showers, the atmosphere is less dense in the same step length (distance to next interaction). The energy loss due to collisions in low density is smaller and acceleration in the electric field is thus more efficient. The effect of the electric field on intensity change is stronger for shower with a higher zenith angle.

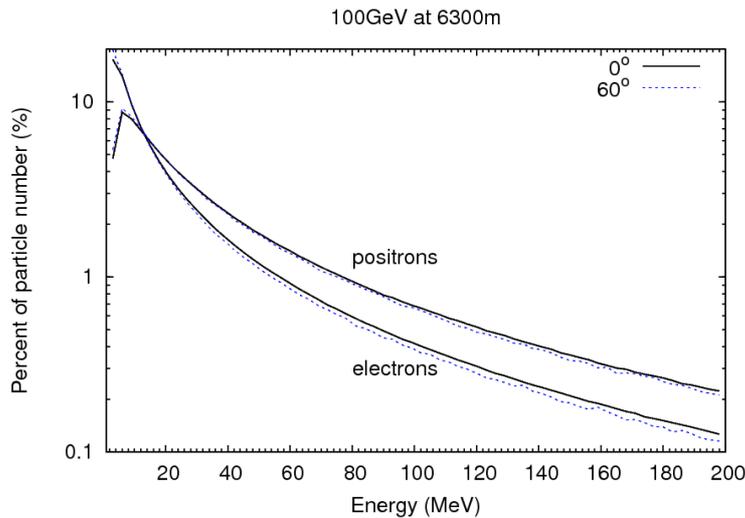

Fig. 14 Percent distributions of electrons and positrons as a function of energy with zenith angles of 0 and 60 degrees in *E* = 0.

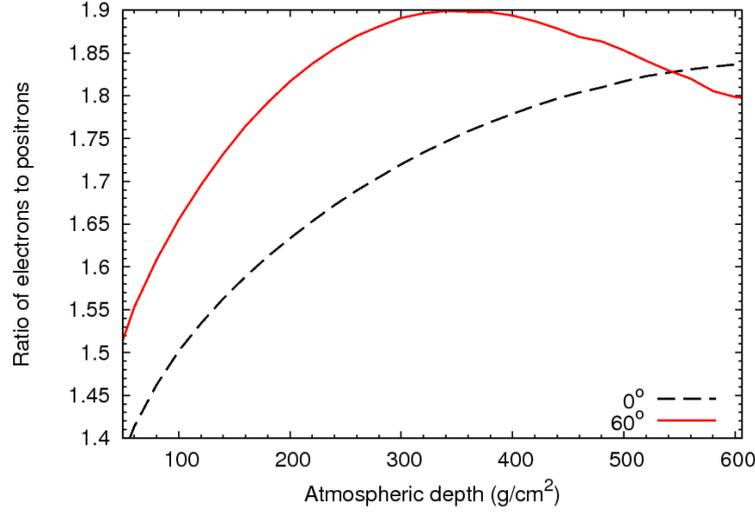

Fig. 15 Ratios of electrons to positrons as a function of atmospheric depth with zenith angles of 0 and 60 degrees in $E = 0$.

**5. Conclusions**

    Monte Carlo simulations are performed with CORSIKA 7.3700 packages to study the intensity change of ground cosmic rays in near-earth thunderstorms electric fields. We conduct simulations for vertical and inclined (with zenith angles of 30 and 60 degrees) proton showers with primary energies of 30, 100 and 770 GeV. The electric fields are chosen as a series of values in the range of -1000−1000 V/cm, and the length of the fields is 2 km from 6300 m to 4300 m.

    The total number of positrons and electrons will increase or decrease in different electric fields. In a negative field, the shower size increases as the field strength increases. The amplitude enhancement is up to 66.5% in -1000 V/cm. In the positive field, the number of positrons continues to increase while the number of electrons continues to decrease. If the positive field is greater than 600 V/cm, the positrons outnumber the electrons at YBJ, and the shower size increases as the field strength increases. The amplitude enhancement is about 14.3% in 1000 V/cm. If the positive field is less than 600 V/cm, the number of electrons remains more than that of positrons. The shower size decreases due to the increase in the number of positrons being less than the decrease in the number of electrons. A certain degree of decline (3.1%) occurs at YBJ in our simulations. The decreases also occur in different primary energy showers and inclined showers. There are two main factors that may be considered for the decreasing phenomenon. First, the number of positrons is less than that of electrons because of the asymmetry of mechanism. Secondly, the electric field has more obvious effects on electrons which have smaller energies than that of positrons.

    In 2002, Alexeenko et al. reported the declines for soft component of cosmic rays in positive fields. The ARGO-YBJ experiment also detected a few events with decreasing phenomena during thunderstorms. The declining phenomena in our simulation are consistent with the ground-based experimental observations above. Accordingly, we believe the intensity decline of ground cosmic

rays, which was detected in mountain top experiments, is really related to the near-earth thunderstorms electric field. Our simulation results are useful in understanding the decreasing phenomenon and give more information about the acceleration mechanism.

*This paper is supported by the National Natural Science Foundation of China (Grant No. 11475141). We wish to express our sincere thanks to Prof. Zhu Qingqi, Ding Linkai, Zha Min and Wu Chaoyong.*

**References**
[1]T. C. Marshall, W. D. Rust and M. Stolzenburg, Electrical structure and updraft speeds in thunderstorms over the southern Great Plains, *J. Geophys. Res.*, **100** (1995) 1001-1015.
[2] M. Stolzenburg, T. C. Marshall, W. D. Rust, E. Bruning, D. R. MacGorman and T. Hamlin, Electric field values observed near lightning flash initiations, *Geophys. Res. Lett.* **34** (2007) L04804.
**DOI:** 10.1029/2006GL028777
[3] T. C. Marshall, M. Stolzenburg, C. R. Maggio and L. M. Coleman, Observed electric fields associated with lightning initiation, *Geophys. Res. Lett.* **32** (2005) L03813.  **DOI:** 10.1029/2004GL021802
[4] C. T. R. Wilson, The electric field of a thundercloud and some of its effects, *Proc. Phys. Soc.,* London, **37** (1924) 32D-37D.
[5] A. V. Gurevich, G. M. Milikh and R. Roussel-Dupre, Runaway electron mechanism of air breakdown and preconditioning during a thunderstorm, *Phys. Lett. A,* **165** (1992) 463-468. DOI: 10.1016/0375-9601(92)90348-P
[6] L. P. Babich, I. M. Kutsyk, E. N. Donskoy, and A.Y. Kudryavtsev, New data on space and time scales of relativistic runaway electron avalanche for thunderstorm environment: Monte Carlo calculations, *Phys. Lett. A*, **245** (1998) 460-470. DOI: 10.1016/S0375-9601(98)00268-0
[7] T. C. Marshall, W. Rison, W. D. Rust, M. Stolzenburg, J. C. Willett and W. P. Winn, Rocket and balloon observations of electric field in two thunderstorms, *J. Geophys. Res.*, **100** (1995) 20815-20828.
DOI: 10.1029/95JD01877
[8] J. R. Dwyer, A fundamental limit on electric fields in air, *Geophys. Res. Lett.* **30** (2003) 2055.
DOI: 10.1029/2003GL017781
[9] E. M. D. Symbalisty, R. Roussel-Dupre, V. A. Yukhimuk, Finite volume solution of the relativistic Boltzmann equation for electron avalanche studies, *IEEE Trans. Plasma Sci.*, **26** (1998) 1575-1582. DOI: 10.1109/27.736065
[10] V. V. Alexeyenko, A. E. Chudakov, V. G. Sborshikov and V. A. Tizengauzen, Short perturbations of cosmic ray intensity and electric field in atmosphere, *Proc. 19th ICRC*, La Jolla, USA, August 11-23, **5**(1985) p352-355.
[11] V. V. Alexeenko, N.S. Khaerdinov, A.S. Lidvansky and V.B.Petkov, Transient variations of secondary cosmic rays due to atmospheric electric field and evidence for pre-lightning particle acceleration, *Phys. Lett. A*, **301** (2002) 299-306. DOI: 10.1016/S0375-9601(02)00981-7
[12] S. Vernetto for EAS-TOP Collaboration, The EAS counting rate during thunderstorms. *Proc. of 27th ICRC*, Copernicus Gesellschaft, Hamburg, Germany, August 7-15, **10** (2001) p4165-4168.
[13] A. Chilingarian, A. Daryan, K. Arakelyan et al. Ground-based observations of thunderstorm correlated fluxes of high-energy electrons, gamma rays, and neutrons, *Phys. Rev. D*, **82** (2010) 043009.
DOI: http://dx.doi.org/10.1103/PhysRevD.82.043009
[14] A. Chilingarian G. Hovsepyan, A. Hovhannisyan, Particle bursts from thunderclouds: Natural particle accelerators above our heads, *Phys. Rev. D*, **83** (2011) 062001. DOI: 10.1103/PhysRevD.83.062001
[15] A. Chilingarian, N. Bostanjyan, L. Vanyan, Neutron bursts associated with thunderstorms, *Phys. Rev. D*, **85** (2012) 085017. DOI: http://dx.doi.org/10.1103/PhysRevD.85.085017
[16] A. Chilingarian, Thunderstorm ground enhancements-Model and relation to lighting flashes, *Journal of Atmospheric and Solar-Terrestrial Physics*, **107** (2014) 68-76.
DOI: 10.1016/j.jastp.2013.11.004
[17] A. Chilingarian, B. Mailyan and L. Vanyan, Recovering of the energy spectra of electrons and gamma rays coming from the thunderclouds, *Atmos. Res.*, **114-115** (2012) 1-16. DOI:10.1016/j.atmosres.2012.05.008

[18] H. Tsuchiya, K. Hibino and K. Kawata et al., Observation of thundercloud-related gamma rays and neutrons in Tibet, *Phys. Rev. D*, **85** (2012) 092006. DOI: 10.1103/PhysRevD.85.092006

[19] H. Tsuchiya, T. Enoto, S. Yamada et al., Detection of high-energy gamma rays from winter thunderclouds, *Phys. Rev. Lett.*, **99** (2007) 165002. DOI: 10.1103/PhysRevLett.99.165002

[20] H. Tsuchiya T. Enoto, T. Torii et al., Observation of an energetic radiation burst from mountain-top thunderclouds, *Phys. Rev. Lett.*, **102** (2009) 255003. DOI:http://dx.doi.org/10.1103/PhysRevLett.102.255003

[21] T. Torii, M. Takeishi, T. Hosono, Observation of gamma‐ray dose increase associated with winter thunderstorm and lightning activity, *J. Geophys. Res.*, **107** (2002) 4324-4332. DOI: 10.1029/2001JD000938

[22] T. Torii, T. Sugita, S. Tanabe, Y. Kimura, M. Kamogawa, K. Yajima, H. Yasuda, Gradual increase of energetic radiation associated with thunderstorm activity at the top of Mt. Fuji, *Geophys. Res. Lett.*, **36** (2009) L13804. DOI:10.1029/2008GL037105

[23] F. Fuschino, M. Marisaldi and C. Labanti et al., High spatial resolution correlation of AGILE TGFs and global lightning activity above the equatorial belt, *Geophys. Res. Lett.*, **38** (2011) L14806. DOI: 10.1029/2011GL047817

[24] M. S. Briggs, G. J. Fishman and V. Connaughton et al., First results on terrestrial gamma ray flashes from the Fermi Gamma-ray Burst Monitor, *J. Geophys. Res.*, **115** (2010) A07323. DOI: 10.1029/2009JA015242

[25] J. R. Dwyer and M. A. Uman, The Physics of Lightning, *Physics Reports*, **534** (2014) 147-241. DOI:10.1016/j.physrep.2013.09.004

[26] P. Schellart, T. N. G. Trinh, S. Buitink et al., Probing Atmospheric Electric Fields in Thunderstorms through Radio Emission from Cosmic-Ray-Induced Air Showers, *Phys. Rev. Lett.*, **114** (2015) 165001. DOI:http://dx.doi.org/10.1103/PhysRevLett.114.165001

[27] A. Dubinova, C. Rutjes, U. Ebert, S. Buitink, O. Scholten and G.T.N. Trinh, Prediction of Lightning Inception by Large Ice Particles and Extensive Air Showers, *Phys. Rev. Lett.*, **115** (2015) 015002. DOI:http://dx.doi.org/10.1103/PhysRevLett.115.015002

[28] A.V. Gurevich, and A. N. Karashtin, Runaway Breakdown and Hydrometeors in Lightning Initiation, *Phys. Rev. Lett.*, **110** (2013) 185005. DOI:http://dx.doi.org/10.1103/PhysRevLett.110.185005

[29] A. Chilingarian, G. Hovsepyan, G. Khanikyanc, A. Reymers and S. Soghomonyan, Lighting origination and thunderstorm ground enhancements terminated by the lightning flash, *EPL*, **110** (2015) 49001. DOI: 10.1209/0295-5075/110/49001

[30] E. S. Cramer, J. R. Dwyer, S. Arabshahi, I. B. Vodopiyanov, N. Liu and H. K. Rassoul, An analytical approach for calculating energy spectra of relativistic runaway electron avalanches in air, *J. Geophys. Res. Space Physics*, **119** (2014) 7794-7823. DOI: 10.1002/2014JA020265

[31] J. R. Dwyer, Implications of x-ray emission from lightning, *Geophys. Res. Lett.*, **31** (2004) L12102. DOI:10.1029/2004GL019795

[32] L. P. Babich, E. I. Bochkov, E. N. Donskoi and I. M. Kutsyk, Source of prolonged bursts of high-energy gamma rays detected in thunderstorm atmosphere in Japan at the coastal area of the Sea of Japan and on high mountaintop, *J. Geophys. Res.*, **115** (2010) A09317. DOI: 10.1029/2009JA015017

[33] S. Buitink, T. Huege, H. Falcke, D. Heck and J. Kuijpers, Monte Carlo simulations of air showers in atmospheric electric fields, *Astropart. Phys.*, **33** (2010) 1-12. DOI:10.1016/j.astropartphys.2009.10.006

[34] M. Tavani, M. Marisaldi, C. Labanti et al., Terrestrial Gamma-Ray Flashes as Powerful Particle Accelerators, *Phys. Rev. Lett.* **106** (2011) 018501. DOI: http://dx.doi.org/10.1103/PhysRevLett.106.018501

[35] X. M. Zhou, N. Ye, F. R. Zhu et al., Observing the effect of the atmospheric electric field inside thunderstorms on the EAS with the ARGO-YBJ experiment, Proc. of *32nd ICRC*, Beijing, China, August 11-18, **11** (2011) p287-290.

[36] Y. Zeng, F. R. Zhu, H. Y. Jia et al., Correlation between cosmic ray flux and electric atmospheric field variations with the ARGO-YBJ experiment, Proc. of *33rd ICRC*, Rio de Janeiro, Brazil, July 2-9, 2013, p0757-0760

[37] D. Heck and J. Knapp et al., CORSIKA: A Monte Carlo Code to Simulate Extensive Air Showers, **FZKA**: Forschungszentrum Karlsruhe GmbH, Karlsruhe, 1998, Vol. **6019**

Available from: http://www-ik.fzk.de/corsika/physics_description/corsika_phys.html

[38] A. F. Bielajew, *Electron Transport in **E** and **B** Fields, in Monte Carlo Transport of Electrons and Photons* (New York: Plenum Press), 1988, pp. 421-434.

[39] D. X. Liu, X. S. Qie, Z. C. Wang, X. K. Wu, and L. X. Pan, Characteristics of lightning radiation source distribution and charge structure of squall line. *Acta Phys. Sin.*, **62** (2013) 219201. DOI: 10.7498/aps.62.219201

[40] X. X. Zhou, X. J. Wang, D. H. Huang, H. Y. Jia and C. Y. Wu, Simulation study on the correlation between the ground cosmic rays and the near earth thunderstorms electric field at Yangbajing (Tibet China), *Acta Phys. Sin.*, **64** (2015) 149202. DOI: 10.7498/aps.64.149202

[41] G. Aielli for ARGO-YBJ Collaboration, Search For Gamma Ray Bursts With the ARGO-YBJ Detector in Scaler Mode, *Astrophys. J.*, **699** (2009) 1281-1287. DOI: 10.1088/0004-637X/699/2/1281

[42] H. A. Bethe, Theory of passage of swift corpuscular rays through matter, *Ann. Phys.*, **5** (1930) 325-400.

[43] R. Roussel-Dupre and A. V. Gurevich, On runaway breakdown and upward propagating discharges, *J. Geophys. Res.*, **101** (1996) 2297-2311. DOI: 10.1029/95JA03278